\documentclass[prl,twocolumn,superscriptaddress,citeautoscript,amsart,longbibliography,titlepage]{revtex4-2}

\usepackage[labelfont=bf,font=small]{caption}
\usepackage{graphicx}
\usepackage{multirow}
\usepackage{color}
\usepackage{bm}
\usepackage{times}
\usepackage{amsmath,bm,amsfonts}
\usepackage{dcolumn}
\usepackage{graphicx}
\usepackage{latexsym}
\usepackage{hhline}
\usepackage{comment}

\usepackage{braket}
\usepackage{bbold}
\usepackage{multirow}

\def\ket#1{|#1\rangle }

\def\d{\partial}

\begin{document}
\title{Circular-polarization-selective perfect reflection from chiral superconductors}

\author{Junyeong \surname{Ahn}}
\email{junyeong.ahn@austin.utexas.edu}
\affiliation{Department of Physics, Harvard University, Cambridge, MA 02138, USA}
\affiliation{Department of Physics, The University of Texas at Austin, Austin, TX 78712, USA}

\author{Ashvin \surname{Vishwanath}}
\email{avishwanath@g.harvard.edu}
\affiliation{Department of Physics, Harvard University, Cambridge, MA 02138, USA}

\date{\today}

\begin{abstract}
{\bf
Integrating mirrors with magnetic components is crucial for constructing chiral optical cavities, which provide tunable platforms for time-reversal-asymmetric light-matter interactions.
Here, we introduce single-crystal circular-polarization-selective mirrors based on chiral superconductors, which break time-reversal symmetry themselves, eliminating the need for additional components.
We show that a circular-polarization-selective perfect reflection (CSPR) occurs for strong-coupling superconductors in the BCS-BEC crossover regime or beyond if the optical Hall conductivity is significant in the unit of conductivity quantum per unit layer, $e^2/ha_z$, where $a_z$ is the lattice constant along the surface normal.
While the optical Hall conductivity in chiral superconductors is typically tiny, we classify three routes to obtain a large value.
We demonstrate the significant optical Hall conductivity and the resulting CSPR with two examples:
(1) superconductivity in doped quantum Hall insulators and (2) chiral pairing that preserves the Bogoliubov Fermi surfaces in the weak-pairing limit.
We also discuss the application of our theory to the recently discovered chiral superconducting phase in rhombohedral graphene.
Our theory reveals the potential of these classes of chiral superconductors as promising elements for building high-quality-factor terahertz chiral cavities.
}
\end{abstract}

\maketitle

{\bf \noindent Introduction.}
Optical cavities are emerging as novel platforms to engineer materials properties and chemical reactions as well as qubit states, making their design with high quality an important issue in quantum materials science~\cite{hubener2021engineering,schlawin2022cavity,bhuyan2023rise}.
The quantum confinement of photons between mirrors enhances light-matter coupling, a hallmark of cavity quantum electrodynamics.
Since the cavity photon energy and the coupling strength are tunable by mechanically adjusting the distance between mirrors, specific excitations in matter can be engineered through resonant coupling to cavity photons.
Furthermore, the symmetry of the resulting phases of matter can also be selectively engineered to drive the system into desired target phases or unexplored new phases~\cite{hubener2021engineering}.
Despite this potential, traditional designs often require complex structures involving multiple components, which makes it challenging to construct compact symmetry-breaking cavities with high quality factors.

\begin{figure*}[t!]
\includegraphics[width=\textwidth]{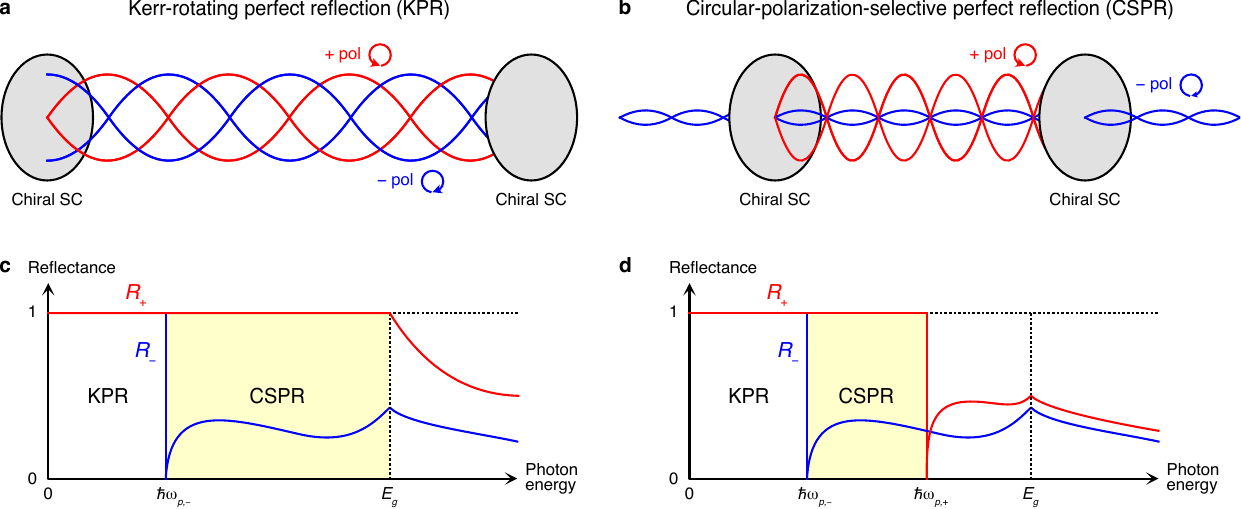}
\caption{
{\bf Chiral perfect reflections in superconductors.}
{\bf a}, Kerr-rotating perfect reflection (KPR).
Both circular polarizations are 100 \% reflected but with different phase shifts.
{\bf b}, Circular-polarization-selective perfect reflection (CSPR).
Only one of the two circular polarizations is reflected 100 \%.
{\bf c} and {\bf d}, Frequency windows for chiral perfect reflections.
Chiral perfect reflections require photon energy below the fermionic excitation gap $E_g$.
KPR occurs for optical frequencies below the lower chiral plasmon energy $\hbar\omega_{p,\rm lower}$, and CSPR occurs between $\hbar\omega_{p,\rm lower}$ and $\min(E_g,\hbar\omega_{p,\rm higher})$.
}
\label{fig:scheme}
\end{figure*}

In this paper, we reveal a unique functional property of superconductors that has the potential to facilitate the design of chiral optical cavities.
We show that chiral superconductors can realize single-crystal mirrors that perfectly reflect one circular polarization and are dielectrics for the other opposite sense of circular polarization, namely, those with circular-polarization-selective perfect reflection (CSPR).
While chirality usually refers to a structural property that lacks any mirror symmetries in three-dimensional space $(x,y,z)$, here we consider the spatio-temporal chirality in three-dimensional spacetime $(t,x,y)$ throughout the paper, meaning the absence of two-dimensional mirror symmetries and time-reversal symmetry.
This type of chirality has the same symmetry as the angular momentum along the $z$ direction. For optical modes, the chirality originates from lifting the degeneracy between two circular polarizations, having different spin angular momenta $l_z=\pm 1$ (i.e., two different circular polarizations), as shown in Figs.~\ref{fig:scheme}a,b.
Thus, our discovery of the circular-polarization-selective perfect reflection provides a new way to build chiral cavities that can be used to engineer spin and orbital magnetism and the associated quantum phases, such as quantum Hall insulators and chiral topological superconductors.
The proposed optical cavities are particularly suited for terahertz frequencies or below.
Since thermal fluctuations significantly reduces the coherence of low-frequency photon modes, cryogenic cooling is crucial for coherent quantum effects, which is the condition ideal for superconducting materials.
Furthermore, as we discuss below, the CSPR frequencies are upper bounded by the size of the superconducting gap, typically smaller than a few terahertz.

The presence of a finite excitation gap is a key physical property of superconductors in our theory, which enables perfect reflections at low frequencies in contrast to ordinary metals.
Building on this feature, our analysis is focused on how to achieve circular-polarization selectivity inside the gap.
We find that the circular-polarization-selective perfect reflection requires (1) strong pair binding where the binding energy is comparable to or larger than the Fermi energy (i.e., the crossover between the weak-coupling Bardeen-Cooper-Schrieffer (BCS) regime and the strong-coupling Bose-Einstein condensate (BEC) regime or beyond) and (2) the significant optical Hall conductivity of an order of the conductivity quantum.
Both conditions are challenging to satisfy.
In particular, the optical Hall conductivity is known to be extremely small in clean chiral superconductors~\cite{roy2008collective,lutchyn2008gauge,taylor2012intrinsic,denys2021origin,yazdani2024polar}.
While there exist a few examples where the Hall conductivity can reach a significant fraction of the conductivity quantum~\cite{can2021probing,song2022doping,zhang2024quantum}, a general understanding of how to generate large Hall effects in superconductors has been lacking.
Here, we provide such a theoretical framework by classifying three possible distinct mechanisms to generate a significant optical Hall conductivity in clean chiral superconductors, employing the selection rules recently developed~\cite{xu2019nonlinear,ahn2021theory,ahn2021many}.
We provide concrete model examples that show large Hall conductivity and the circular-polarization-selective perfect reflection.
Based on our analysis, the chiral superconducting phase appearing in rhombohedral graphene~\cite{han2024signatures} is a promising candidate for observing signatures of circular-polarization-selective perfect reflection, if it can be realized with a layered three-dimensional structure.
Our results motivate and guide the search for candidate strong-coupling superconductors showing circular-polarization-selective perfect reflection.
\\

{\bf \noindent Circular-polarization-selective perfect reflection.}
Electromagnetic responses of materials are described by the electric susceptibility in the linear-response regime, where the light electric field is not so intensive.
The electric susceptibility of superconductors has three contributions:
\begin{align}
\chi^{ab}(\omega)
&
=-\frac{D^{ab}}{\epsilon_0\omega^2}
+\tilde{\chi}_S^{ab}(\omega)
+\frac{\sigma^{ab}_H(\omega)}{-i\omega\epsilon_0},
\end{align}
where $D^{ab}$ is the superfluid weight, $\tilde{\chi}_S^{ab}$ is the symmetric part of the interband electric susceptibility, and $\sigma^{ab}_H=(\sigma^{ab}-\sigma^{ba})/2$ is the optical Hall conductivity (see Eq.~\eqref{eq:conductivity} and Methods).
It is the Hall part that makes light propagation circular-polarization-dependent.

\begin{table}[b!]
\begin{tabular}{c|cc}
Property				&Materials	&Conditions	\\
\hhline{=|==}
$T=0$		&Metals or SCs		&$\omega<\omega_p$\\
$A=0$			&Insulators or SCs	&$\hbar\omega<E_g$\\
$R_+\ne R_-$	&Magnets or Chiral SCs		&$\sigma_H\ne 0$ ($\omega_{p,-}\ne \omega_{p,+}$)\\
CSPR			&Chiral SCs		&$\omega_{p,\mp }<\omega<\omega_{p,\pm}$ \& $\hbar\omega<E_g$
\end{tabular}
\caption{
{\bf Conditions for circular-polarization-selective perfect reflection (CSPR).}
$T$, $R$, and $A$ are transmittance, reflectance, absorptance, satisfying $T+R+A=1$.
Subscripts $+$ and $-$ denote two different circular polarizations.
SC indicate superconductor, $\omega$ is the optical frequency, $E_g$ is the excitation gap, $\sigma_H(\omega)$ is the optical Hall conductivity, and $\omega_{p,\pm}$ is the plasmon frequency for $\pm$ polarizations.}
\label{tab:CSPR}
\end{table}

Let us analyze the conditions under which circular-polarization-selective perfect reflection occurs.
For perfect reflection, we need zero absorption and zero transmission.
In this regard, the presence of an excitation gap in superconductors is the key advantage of superconductors, which distinguishes them from ordinary metals that absorb light at arbitrary low frequencies and generate Ohmic losses.
Since no absorption has to occur for perfect reflection, we focus on photon energies below the superconducting gap $E_g$, where no fermionic excitations occur such that $\chi^{ab}$ is Hermitian (i.e., ${\rm Im}\tilde{\chi}_S^{ab}=0$ and ${\rm Im}\sigma_H^{ab}=0$).
Accordingly, the refractive indices $n_{\pm}$ are either real or purely imaginary because $n_{\pm}^2$ are real valued as eigenvalues of a Hermitian matrix $\epsilon^{ab}=\epsilon_0(\delta^{ab}+\chi^{ab})$, where $\hat{\pm}$ indicate two circular polarizations.
As the propagation of light inside a medium requires a nonzero real part of the refractive index, no transmission is allowed in the bulk medium when the refractive index is purely imaginary. Therefore, a circular-polarization-selective perfect reflection occurs when one polarization has a pure-imaginary refractive index while the other has a nonzero real component in the refractive index.
We summarize the necessary conditions in Table.~\ref{tab:CSPR}, highlighting chiral superconductors as ideal materials to achieve circular-polarization-selective perfect reflection.

For a more quantitative but simple analysis, let us consider reflections at normal incidence along $\hat{z}$ and assume that the medium has a $C_{3z}$ or $C_{4z}$ rotational symmetry.
Under these conditions, the refractive indices for circular polarizations $\hat{\pm}=(\hat{x}\pm i\hat{y})/\sqrt{2}$ are $n_{\pm}=\sqrt{1+\chi_S\pm i\chi_H}$, where $\chi_S=\chi^{xx}$, and $\chi_H=\chi^{xy}$.
Since we are interested in photon energies smaller than $E_g$, we can approximate $\tilde{\chi}_S$ and $\sigma_H$ with their zero-frequency values, which allows us to analytically calculate the chiral plasmon frequencies by solving quadratic equations $n_{\pm}^2(\omega_{p,\pm})=0$.
We find
\begin{align}
\label{eq:plasma-frequency}
\omega_{p,\pm}
&\approx \sqrt{\frac{D}{\tilde{\epsilon}}+\left(\frac{\sigma_H(0)}{2\tilde{\epsilon}}\right)^2}\pm \frac{\sigma_H(0)}{2\tilde{\epsilon}},
\end{align}
where $D=D^{xx}$, and $\tilde{\epsilon}= \epsilon_0(1+\tilde{\chi}_S(0))$.
When the $\sigma_H(0)$ term dominates over the $D$ term, this expression for the lower plasmon frequency reduces to $\omega_{p,\rm lower}\approx D/|\sigma_H(0)|$.
At small frequencies $\omega<\omega_{p,\pm}$, both circular polarizations are fully reflected, but with a phase difference (equivalently, the Kerr rotation for linear polarizations --- the formula for the Kerr angle is given in a later section below) when the Hall conductivity is finite.
The circular-polarization-selective perfect reflection can occur for $\omega_{p,\rm lower}<\omega<\omega_{p,\rm higher}$, and this frequency window is proportional to the Hall conductivity if both plasmon frequencies are below the superconducting gap: $\delta\omega_p\equiv \omega_{p,+}-\omega_{p,-}=\sigma_H(0)/\tilde{\epsilon}$.
When the higher plasmon frequency is larger than the superconducting gap, circular-polarization-selective perfect reflection occurs for $\hbar\omega_{p,\rm lower}<\hbar\omega<E_g$.
Then, the polarization corresponding to the plasma frequency $\omega_{p,\rm lower}$ is partially reflected and partially transmitted while the one corresponding to $\omega_{p,\rm higher}$ is fully reflected.

In any case, we need $\hbar\omega_{p,\rm lower}<E_g$ for circular-polarization-selective perfect reflection. This condition requires a strong pairing with a gap $E_g$ comparable to or larger than the Fermi energy $E_F=\hbar\omega_F$.
To see this, let us express the superfluid weight in terms of the Fermi energy by considering a layered material with interlayer distance $a_z$ hosting a cylindrical Fermi surface with degeneracy $n_s$, in which case we have $D=n_s\omega_F e^2/2\pi a_z$.
Then, the gap-to-Fermi-energy ratio is bounded from below by (see Methods)
\begin{align}
\label{eq:bound}
\frac{E_g}{E_F}
>n_s\left(\frac{E_g}{e^2/2\pi\tilde{\epsilon}a_z}+\frac{|\sigma_H(0)|}{e^2/ha_z}\right)^{-1}.
\end{align}
When the Hall conductivity is negligible, this lower bound is much larger than one because $E_g\ll e^2/2\pi \tilde{\epsilon}a_z$ for a reasonable pairing gap size, smaller than a few tens of meV.
On the other hand, for a moderate value of the Hall conductivity in the unit of $e^2/ha_z$, a strong pairing is still required, but in the BCS-BEC crossover regime ($E_g\sim E_F$) and not necessarily the BEC regime ($E_g\gg E_F$).

A significant Hall conductivity comparable to or larger than $e^2/ha_z$ is thus favorable for realizing circular-polarization-selective perfect reflection, benefiting from both broader bandwidth and a weaker pairing threshold.
However, most chiral superconductors show tiny Hall conductivity (e.g. $\sim 10^{-4}e^2/ha_z$ at low frequencies in models of SrRuO$_4$~\cite{taylor2012intrinsic,denys2021origin,yazdani2024polar}) with few exceptions~\cite{can2021probing,song2022doping,zhang2024quantum}, and it is not well understood how to generate a large Hall conductivity for superconductors.
In the following, we discuss the origin of the small Hall conductivity in chiral superconductors and provide ways to generate significant Hall conductivity.\\

{\bf \noindent Optical conductivity in superconductors.}
Let us consider the mean-field Hamiltonian of superconductors
\begin{align}
\hat{H}
&=\frac{1}{2}
\sum_{\bf k}
\hat{\Psi}^{\dagger}_{\bf k}H_{\rm BdG}({\bf k})\hat{\Psi}_{\bf k},\notag\\
\label{eq:BdG}
H_{\rm BdG}({\bf k})
&=
\begin{pmatrix}
h({\bf k})&\Delta({\bf k})\\
\Delta^{\dagger}({\bf k})& -h^T(-{\bf k})
\end{pmatrix},
\end{align}
where $\hat{\Psi}_{\bf k}=(\hat{c}_{\bf k},\hat{c}^{\dagger}_{-\bf k})^T$, and $\Delta({\bf k})=-\Delta^T(-{\bf k})$ is the superconducting order parameter.
This Hamiltonian has symmetry $\hat{C}\hat{H}\hat{C}^{-1}=\hat{H}$ under unitary particle-hole conjugation $\hat{C}\hat{c}_{\bf k}\hat{C}^{-1}=\hat{c}_{-\bf k}^{\dagger}$, and $\hat{C}^2=1$.
Equivalently, the Bogoliubov-de Gennes (BdG) Hamiltonian in Eq.~\eqref{eq:BdG} satisfies $CH_{\rm BdG}({\bf k})C=-H_{\rm BdG}(-{\bf k})$ under anti-unitary particle-hole conjugation $C=\tau_x K$, where $\tau_{i=x,y,z}$ is the Pauli matrix.
In the absence of spin-orbit coupling, we can take only one spin species $\hat{\Psi}_{\uparrow\bf k}=(\hat{c}_{\uparrow\bf k},\hat{c}^{\dagger}_{\downarrow,-\bf k})$, whose BdG Hamiltonian is symmetric under an effective particle-hole conjugation defined by the combination of particle-hole conjugation and spin rotation $C_{\rm eff}=i\tau_y K$, where $\Delta=\Delta_{\uparrow\downarrow}$ and $h=h_{\uparrow\uparrow}=h_{\downarrow\downarrow}$.
We drop the subscript ``$\rm eff$" from the notation of the effective particle-hole conjugation in the following.

\begin{figure*}[t!]
\includegraphics[width=\textwidth]{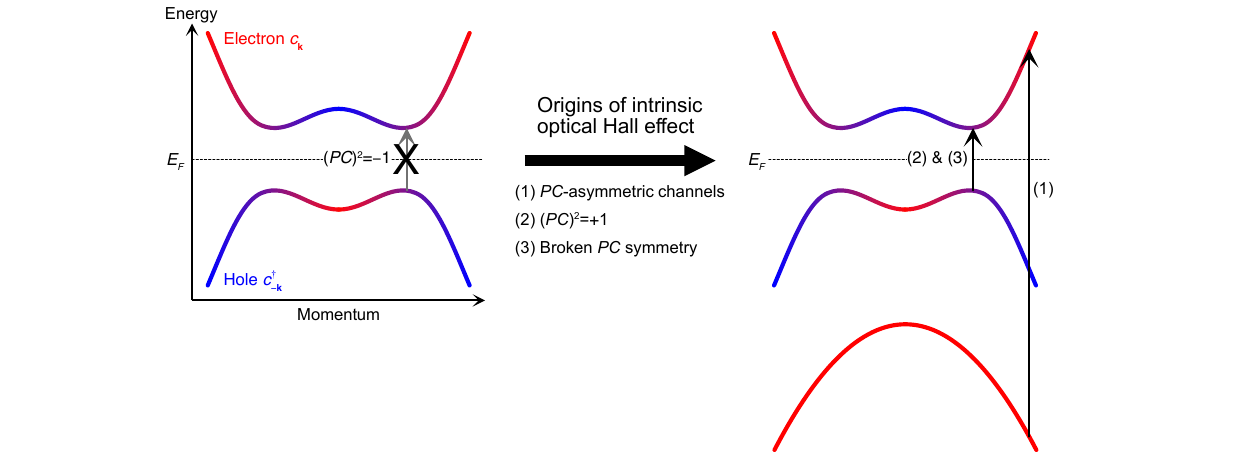}
\caption{
{\bf Origins of the intrinsic optical Hall effect in crystalline chiral superconductors.}
The $PC$-symmetric optical transitions are forbidden in superconductors when $(PC)^2=-1$, which strongly suppresses the anomalous Hall effect, originating from interband (virtual or resonant) optical transitions.
The Hall effect originates from the channels that are allowed by the selection rule.
Here, we focus on intrinsic mechanisms and do not consider extrinsic impurity contributions to the anomalous Hall conductivity.
}
\label{fig:selection-rule}
\end{figure*}

In this work, our analysis focuses on the intrinsic mechanisms of the Hall conductivity, which are dominant mechanisms in clean systems.
In the clean limit, optical conductivity is given by (see Methods)
\begin{align}
\label{eq:conductivity}
\sigma^{ab}(\omega)
&=\frac{i}{\omega}D^{ab}
+\frac{i}{2\hbar}
\int_{\bf k}\sum_{n,m}\frac{f_{nm}J^a_{nm}J^b_{mn}}{(\omega-\omega_{mn})\omega_{mn}},
\end{align}
where $\int_{\bf k}=\int d^dk/(2\pi)^d$ in $d$ spatial dimensions (we consider $d=3$ unless otherwise specified), $\hat{H}\ket{\psi_n}=\hbar\omega_n\ket{\psi_n}$, $f_{nm}=f_{n}-f_{m}$ is the difference of the Fermi distributions, $J^a_{mn}=\braket{\psi_m|\hat{J}^a|\psi_n}$ is the matrix element of the current operator $\hat{\bf J}=-\d \hat{H}/\d {\bf A}|_{{\bf A}=0}$, and $\omega_{mn}=\omega_m-\omega_n$, and $D^{ab}=(\hbar/2)\int_{\bf k}\sum_nf_n\d_{A_a}\d_{A_b}\omega_n|_{{\bf A}=0}$ is the superfluid weight.
The interband matrix elements of the current operator are constrained by symmetries, giving selection rules for optical transitions.\\

{\bf \noindent Optical selection rules.}
Inversion symmetry plays an important role in optical transitions in superconductors when combined with particle-hole symmetry.
For excitations in ordinary metals or insulators, inversion symmetry imposes the selection rule only at special momenta that are invariant under flipping momentum.
However, the combination of particle-hole conjugation $C$ and inversion $P$ gives an additional constraint at every momentum ${\bf k}$~\cite{ahn2021theory,ahn2021many}:
\begin{align}
\label{eq:selection-rule}
\braket{\hat{P}\hat{C}\psi_{n\bf k}|\hat{\bf J}|\psi_{n\bf k}}=0
\text{ for }(\hat{P}\hat{C})^2=-1,
\end{align}
which follows from $\hat{P}\hat{C}\hat{\bf J}(\hat{P}\hat{C})^{-1}=\hat{\bf J}$ and the anti-unitary nature of the single-particle representation for $\hat{P}\hat{C}$.
In the presence of inversion symmetry, $\ket{\psi_{n'\bf k}}\equiv \hat{P}\hat{C}\ket{\psi_{n\bf k}}$ is an energy eigenstate with energy $E_{n'\bf k}=-E_{n\bf k}$.
Therefore, equation~\eqref{eq:selection-rule} imposes an optical selection rule on the transition between two $PC$-related energy eigenstates.
Although bare spatial inversion and particle-hole conjugation satisfy $(PC)^2=P^2C^2=(+1)(+1)=1$, most superconductors have  $P_{\rm eff}C_{\rm eff}$ symmetry with $(P_{\rm eff}C_{\rm eff})^2=-1$, where $\hat{O}_{\rm eff}$ is a combination of $\hat{O}$ with a local unitary operation.
The condition for the latter is satisfied either by (1) odd-parity pairing where effective inversion is the combination of $\pi/2$ phase rotation times spatial inversion, such that $P_{\rm eff}C_{\rm eff}=-C_{\rm eff}P_{\rm eff}$, (2) low-dimensionality where two-fold rotation or mirror plays the role of spatial inversion, $P_{\rm eff}^2=-1$, or (3) spin-singlet pairing in the absence of spin-orbit coupling, where $C_{\rm eff}$ is a combination of particle-hole conjugation and spin flip, such that $C_{\rm eff}^2=-1$.

This selection rule gives important constraints on the Hall conductivity of chiral superconductors, because the Hall response originates from the inter-band term in Eq.~\eqref{eq:conductivity}.
The anomalous Hall effect in superconductors has been studied in the context of topological superconductivity originating from chiral $p+ip$ superconductivity~\cite{roy2008collective,lutchyn2008gauge,taylor2012intrinsic,denys2021origin,yazdani2024polar}.
A chiral pairing $p_x+ip_y$ breaks mirror and time-reversal symmetries, allowing a finite Hall conductivity $\sigma^{xy}_H$.
However, this is not enough for generating a large Hall conductivity
because the chiral gap opening at the Fermi level does not contribute to the Hall response by the selection rule, even when the pairing produces a topological phase.
This situation is very different from the case of Hall responses in insulators, where the Hall conductivity is directly proportional to a topological invariant, the Chern number.

There exist three routes to bypass this selection rule to obtain a large Hall response in superconductors (Fig.~\ref{fig:selection-rule}).
We discuss two examples below that have not been discussed in the literature.

\begin{figure*}[t!]
\includegraphics[width=\textwidth]{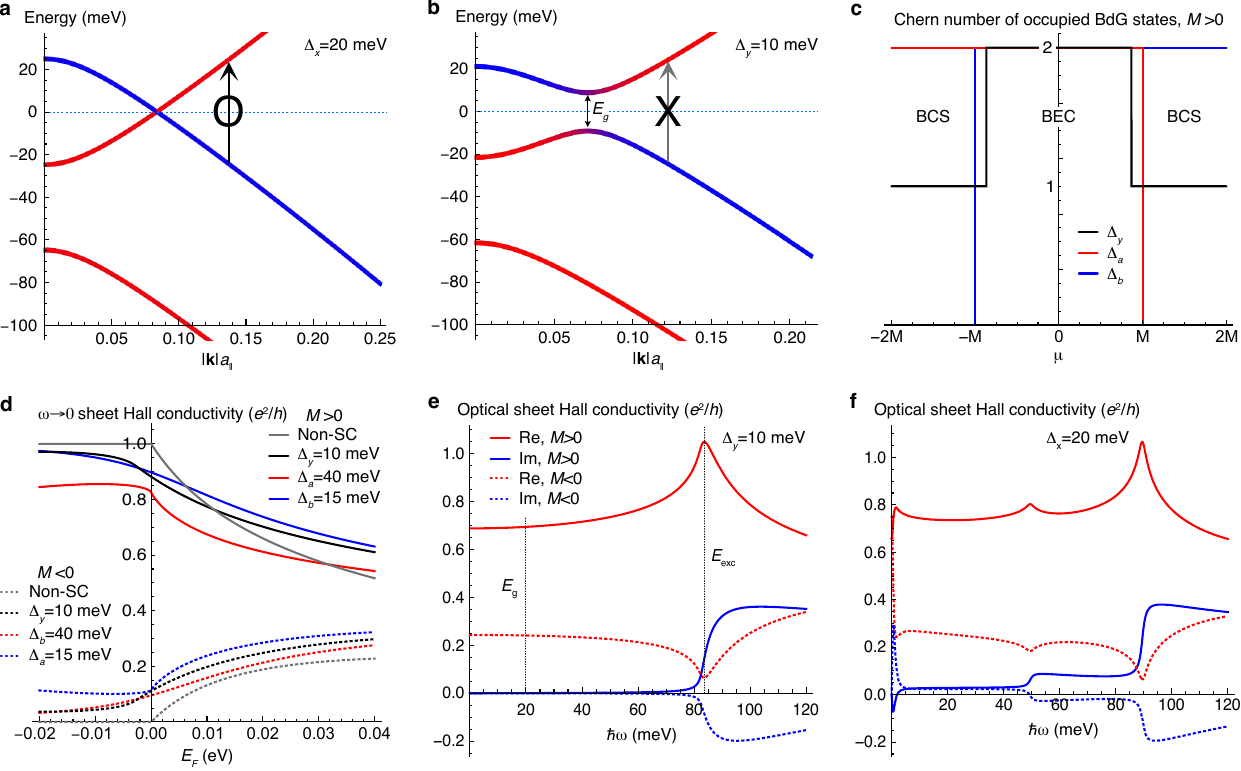}
\caption{
{\bf Optical Hall conductivity in a continuum model of superconductivity from a chiral metal.}
{\bf a} and {\bf b}, Energy bands of the Bogoliubov-de Gennes Hamiltonian.
Line colors indicate the electron $\hat{c}$ (red) vs hole $\hat{c}^{\dagger}$ (blue) characters.
Shown here is the case with $M>0$.
The case with $M<0$ has a similar band structure.
Superconductivity does not open the gap in {\bf a} because the crossing between the electron and hole bands is protected by $PC$ symmetry, where $(PC)^2=1$.
Although the band structure in ${\bf a}$ resembles that of the normal-state metal, the new optical excitation channel relating electron and hole states develops by superconducting pairing $\Delta_x$.
In {\bf b}, where $(PC)^2=-1$, a full gap opens.
Similarly, $\Delta_{a,b}$ pairings also open the full gap (not shown).
In these cases, the $PC$-symmetric optical excitations are forbidden within electric-dipole approximation.
{\bf c}, Chern number of the occupied BdG states for $M>0$.
This number equals the number of left-chiral Majorana edge states minus the number of right-chiral Majorana edge states.
{\bf d}, Hall conductivity in the zero-frequency limit.
$M>0$ for solid and $M<0$ for dashed lines.
Gap functions are taken to produce the superconducting gap of about 20 meV.
{\bf e}, Optical Hall conductivity for $\Delta_y=10$ meV.
$\Delta_{a,b}$ pairing leads to similar spectra.
We take a Lorentzian broadening $\Gamma=1$ meV of resonant transitions.
The optical Hall conductivity develops a finite imaginary value above the optical excitation gap $E_g$ meV.
Resonant transitions are forbidden for $E_g\le \hbar\omega<E_{\rm exc}$ because of the $PC$-symmetry of the BdG Hamiltonian.
For the $\Delta_x$ pairing in {\bf f}, a resonant optical transition occurs down to zero frequency, i.e., $E_g=E_{\rm exc}=0$.
For all plots, we take $\hbar v/a_{\parallel}=A/a_{\parallel}^2=0.5$ eV, $|M|=20$ meV, and $E_F= |\mu|-|M|=20$ meV in Eq.~\eqref{eq:Dirac-model} and consider superconducting pairing of Eq.~\eqref{eq:Dirac-model-SCgap}.
}
\label{fig:Dirac-model}
\end{figure*}
{\bf \noindent Superconductivity from chiral metals.}
The first route uses the $PC$-asymmetric transition channels.
In particular, we propose harnessing the Hall response existing in the normal state, which does not rely on the chiral superconducting pairing.
The primary role of superconductivity here is to open the optical excitation gap, which is important for applications as a mirror.

We consider a two-band model of layered two-dimensional chiral metal.
For simplicity, we neglect the dispersion along the stacking direction $z$ and treat it as a two-dimensional model.
\begin{align}
\label{eq:Dirac-model}
h({\bf k})=-\mu+\hbar v(k_x\sigma_x+k_y\sigma_y)+(Ak^2-M)\sigma_z,
\end{align}
where $\sigma_{i=x,y,z}$ are orbital Pauli matrices.
This model has spatial inversion symmetry under $P=\sigma_z$ and emergent continuous rotational symmetry under $R(\theta)=\exp(-i\theta \sigma_z/2)$ but breaks in-plane mirror $M_{x,y}$ and time reversal $T$ symmetries, so it generically has an out-of-plane orbital magnetic moment.
Here, continuous rotational symmetry emerges due to our simple choice of the model, but we note that it is not an exact symmetry of a physical spin-free orbital-based system as it has spin-like representation $R(2\pi)=-1$.
In this model, we have a quantum Hall insulator ($c_1=1$) for $M>0$ and a topologically trivial insulator ($c_1=0$) for $M<0$ at half filling, where $c_1$  is the Chern number of the occupied states.
The Chern number of the occupied BdG bands is $c_1^{\rm BdG}=2c_1$ because we double the degrees of freedom to include both electrons and holes.

For superconductivity, we include the constant pairing
\begin{align}
\label{eq:Dirac-model-SCgap}
\Delta({\bf k})
=\begin{pmatrix}
\Delta_a&\Delta_x-i\Delta_y\\
\Delta_x+i\Delta_y&\Delta_b
\end{pmatrix}.
\end{align}
Orbital-selective pairings for the $\Delta_{a}$ and $\Delta_{b}$ channels have even parity and angular momentum $l=1$ and $l=-1$ under $R(\theta)$, respectively, and the inter-orbital pairings $\Delta_{x}$ and $\Delta_{y}$ channels have odd parity and $l=0$.
Here, $\Delta_x$ does not open the gap for small pairing because $PC=\tau_xK$ symmetry of the BdG Hamiltonian, with $(PC)^2=1$, protects the Bogoliubov Fermi surface~\cite{agterberg2017bogoliubov,bzduvsek2017robust} [Fig.~\ref{fig:Dirac-model}a].
The others can open the full gap on the Fermi surface because they are symmetric under $PC=i\tau_yK$, satisfying $(PC)^2=-1$ [Fig.~\ref{fig:Dirac-model}b].
The superconducting gap for each pairing channel is isotropic and takes a simple form when $A=0$:
\begin{align}
E_{\Delta_a}
&=
2|\Delta_{a}|\sqrt{(\mu-M)^2/(|\Delta_{a}|^2+(2\mu)^2)},\notag\\
E_{\Delta_b}&=
2|\Delta_{b}|\sqrt{(\mu+M)^2/(|\Delta_{b}|^2+(2\mu)^2)},\notag\\
E_{\Delta_y}&=
2|\Delta_y|\sqrt{1-(M/\mu)^2},
\end{align}
where $\Delta_{j}=0$ for $j\ne i$ is assumed for pairing channel $i$.
The superconducting gaps vanish when $\mu=M$ or $-M$ because a topological phase transition occurs.
Figure~\ref{fig:Dirac-model}c show the case with $M>0$.
At small $\mu$, the superconducting state is adiabatically connected to the BEC limit where Cooper pairs are tightly bound.
In this phase, the topology is the same as the topology of the normal state with $\mu=0$ (namely, $c_1^{\rm BdG}=2c_1=2$ for $M>0$), regardless of the type of the pairing.
The phases with $|\mu/M|> 1$ is adiabatically connected to the weak-coupling BCS limit where $E_F\gg \Delta$, and their topology is sensitive to the form of pairing.

Our calculation of the Hall conductivity in the zero-frequency limit in Fig.~\ref{fig:Dirac-model}d shows that the Hall conductivity is not dramatically changed by superconductivity despite the differences in topological phases for $\Delta_{y,a,b}$.
Therefore, for small doping, the doped quantum Hall insulator ($M>0$) shows larger Hall effects than the doped trivial insulator ($M<0$) in the superconducting states as well as in the normal metallic state.
We note that the Hall conductivity is not quantized even though the full gap opens for fermionic excitations.
This situation is different from the case of the Hall responses in insulators, where the Hall conductivity arises because of chiral edge states and is thus directly proportional to a topological invariant, the Chern number.

The optical Hall conductivity spectrum in Fig.~\ref{fig:Dirac-model}e is shown only for $\Delta_{y}$ but those with $\Delta_{a,b}$ look similar.
In particular, the optical excitation gap $E_{\rm exc}$, where ${\rm Im}\sigma_H$ becomes nonzero, is mostly determined by the excitation gap in the normal state and is larger than the energy gap $E_g$, because $PC$ symmetry with $(PC)^2=-1$ forbids low-energy optical excitations mixing electron and hole states.
In contrast, the $\Delta_{x}$ pairing allows optical excitations down to zero frequency [Fig.~\ref{fig:Dirac-model}f], since $(PC)^2=1$ protects the Bogoliubov Fermi surface and allows low-frequency optical excitations.\\
\begin{figure*}[t!]
\includegraphics[width=\textwidth]{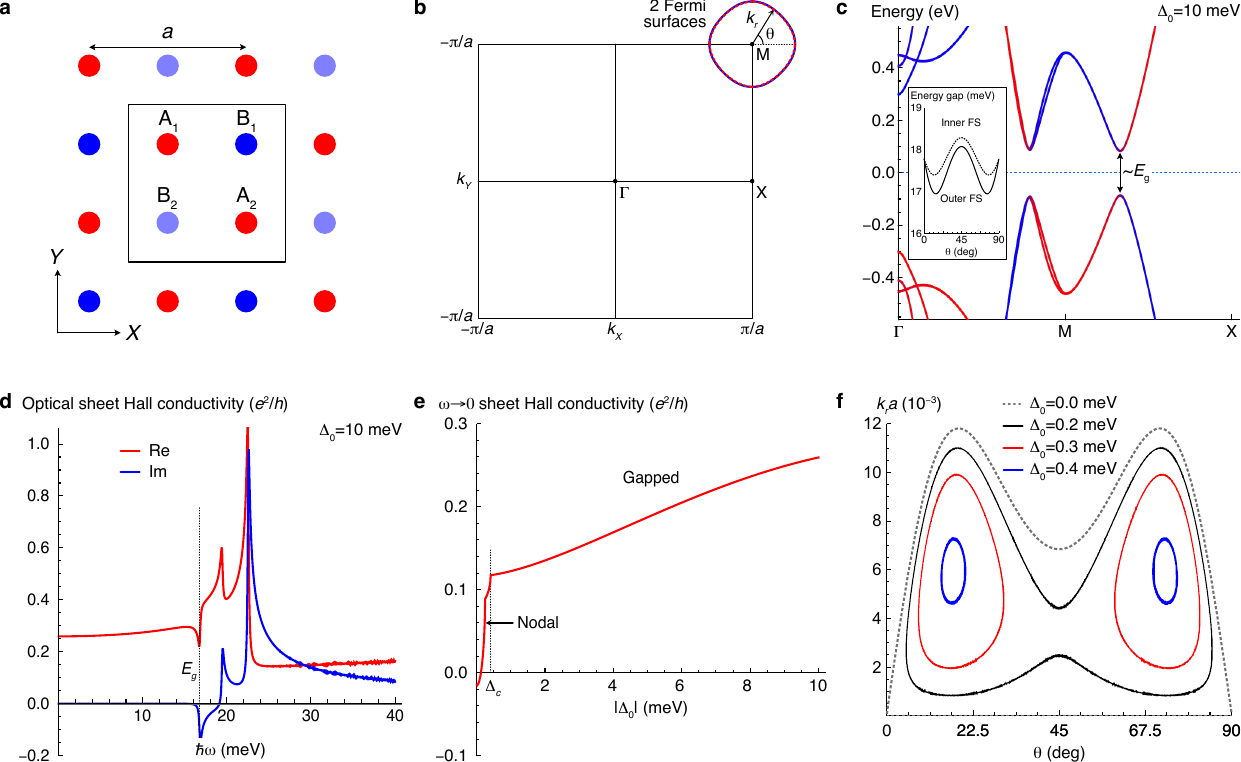}
\caption{
{\bf Optical Hall conductivity in a tight-binding model of $d_{xz}+id_{yz}$-wave superconductors.}
{\bf a}, Lattice structure.
B$_1$ and B$_2$ atoms are displaced oppositely along the out-of-plane direction.
The square box represents the unit cell, and $a$ is the in-plane lattice constant.
We take hopping parameters that are relevant for the electron-doped FeSe.
{\bf b}, Fermi surfaces.
There are two almost overlapping Fermi surfaces near $M$, each with twofold spin degeneracy.
{\bf c}, Energy bands of the Bogoliubov-de Gennes Hamiltonian with a $d_{xz}+id_{yz}$-wave pairing amplitude $\Delta_0$.
Line colors indicate the electron $\hat{c}$ (red) vs hole $\hat{c}^{\dagger}$ (blue) characters.
The inset shows the angular dependence of the superconducting gap on the Fermi surfaces.
{\bf d}, Optical Hall conductivity.
We take a Lorentzian broadening $\Gamma=0.1$ meV.
Unlike in Fig.~\ref{fig:Dirac-model}, the optical excitation gap $E_{\rm exc}$ coincides with the spectral gap $E_g$.
{\bf e}, Pairing dependence of the Hall conductivity in the zero-frequency limit.
{\bf f}, Evolution of the Bogoliubov Fermi surface with varying $\Delta_0$ below $\Delta_c$.
}
\label{fig:FeSe-model}
\end{figure*}

{\bf \noindent Chiral superconducting pairing.}
When the normal state has time reversal symmetry, the Hall conductivity after superconducting phase transition mainly arises from the chiral pairing.
Therefore, to obtain large Hall conductivity, we need nonzero virtual transition amplitudes at the pairing scale by either breaking $P$ symmetry, or preserving $P$ symmetry but realizing $(PC)^2=1$.
The former can be realized by a chiral stacking of two-dimensional materials as in Ref.~\cite{can2021probing,song2022doping}.
Here, we provide an example of the latter, which preserves inversion symmetry.

We consider a five-$d$-orbital tight-binding model on the $A$ sites of the lattice shown in Fig.~\ref{fig:FeSe-model}a.
\begin{align}
\label{eq:FeSe-model}
\hat{h}
=\sum_{i,j}t_{ij}\hat{c}^{\dagger}_{i}\hat{c}_{j},
\end{align}
where $i,j$ are lattice sites, and the spin and orbital indices are implicit.
We take the hopping amplitudes that are relevant for the heavily electron-doped iron selenide and neglect spin-orbit coupling (see Methods).
We also neglect the tunneling between FeSe layers, so that we can consider a two-dimensional Hamiltonian.
This model has symmetries under glide mirror under $G_z:(x,y,z)\rightarrow (x+a/2,y,-z)$ and mirror $M_x:(x,y,z)\rightarrow (-x,y,z)$, $M_{x-y}:(x,y,z)\rightarrow (y,x,z)$, fourfold rotation $C_{4z}:(x,y,z)\rightarrow (y,-x,z)$, where the coordinate center is on a chalcogen atom.
In the normal state, this model hosts two almost degenerate Fermi surfaces enclosing $M=(\pi/a,\pi/a)$ [Fig.~\ref{fig:FeSe-model}b].

We consider a spin-singlet interorbital pairing having $d_{xz}\pm id_{yz}$-like symmetry (i.e., $E_{g}$ irreducible representation) under $D_{4h}$ point group operations.
\begin{align}
\Delta({\bf k})
=2\Delta_0\sigma_y\rho_y\otimes
\begin{pmatrix}
0&0&\cos_+&i\cos_-&0\\
0&0&0&0&0\\
-\cos_+&0&0&0&0\\
-i\cos_-&0&0&0&0\\
0&0&0&0&0
\end{pmatrix},
\end{align}
where $\sigma_y$ and $\rho_{y}$ are Pauli matrices for the spin and sublattice indices, respectively, and $\cos_\pm=\cos[(k_X\pm k_Y)a/2]$.
This pairing opens a full gap with weak angular dependence and induces a chiral topological superconductivity with $c_1^{\rm BdG}=4$ when $\Delta_0$ is sufficiently large [Fig.~\ref{fig:FeSe-model}c].

The optical transition amplitudes with different glide mirror eigenvalues are zero for in-plane-polarized light.
Within each given mirror eigensector, the effective $PC$ operation is $C_{2z}C$, and this symmetry does not forbid optical transitions at low energies because $(C_{2z}C)^2=1$.
Therefore, the optical transitions can occur in the pairing gap scale.

As time reversal and mirror $M_x,M_y$ symmetries are broken by chiral pairing, a finite optical Hall conductivity is generated [Fig.~\ref{fig:FeSe-model}d], and it is sensitive to the size of the pairing amplitude.
Figure~\ref{fig:FeSe-model}e shows that the Hall conductivity in the zero-frequency limit $\omega\rightarrow 0$ is positively correlated with the pairing amplitude.
The pairing dependence has a cusp at $\Delta_c$, above which the superconducting gap fully opens.
For weak pairing, the superconducting gap does not open because the Fermi surfaces carry topological charges for $(PC)^2=1$ and are thus stable under small perturbations until shrunken to a point [Fig.~\ref{fig:FeSe-model}f].\\

\begin{figure*}[t!]
\includegraphics[width=\textwidth]{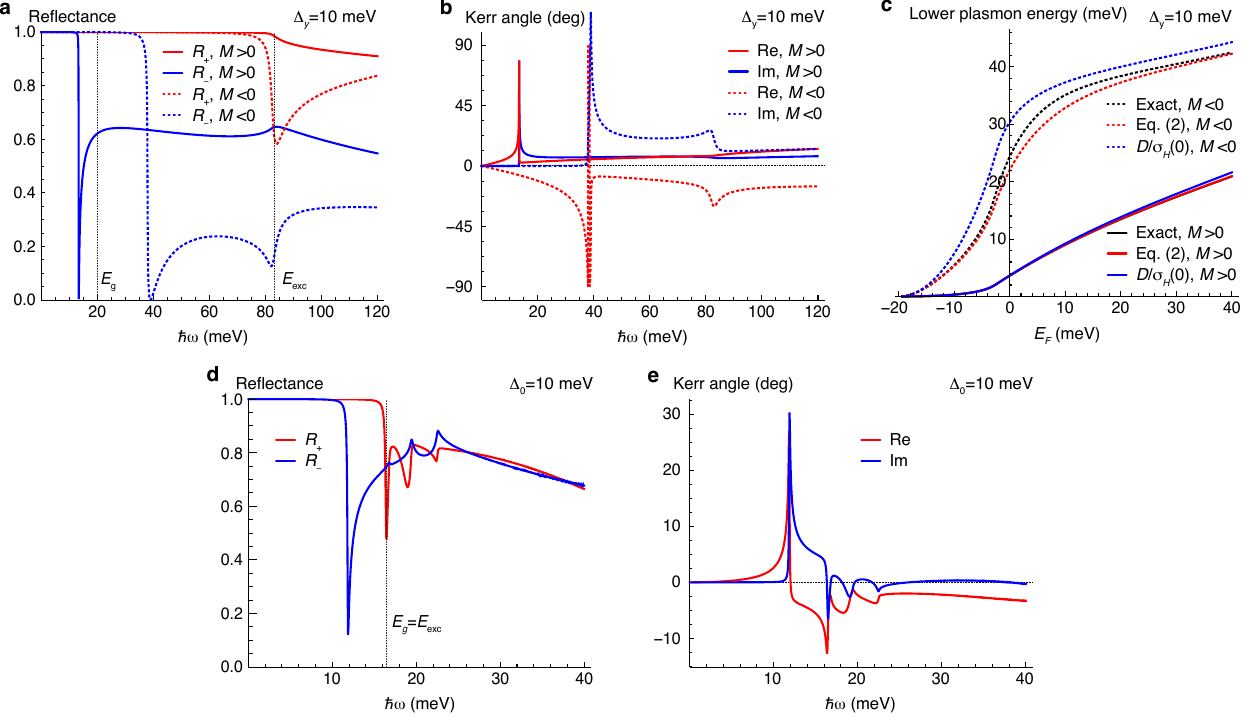}
\caption{
{\bf Magneto-optic Kerr effects in fully gapped chiral superconductors.}
{\bf a}-{\bf c}, Two-band model of superconductivity from a chiral metal with $\Delta_y=10$ meV.
$|M|=20$ meV, $\hbar v/a_{\parallel}=A/a_{\parallel}^2=0.5$ eV, and $\Gamma=1$ meV.
To regularize the UV divergence in calculating the superfluid weight using Eq.~\eqref{eq:SF-weight}, we calculate it with reference to $\mu=0$. Namely, we replace $D^{xx}$ with $D^{xx}-D^{xx}(\mu=0)$.
In {\bf a} and {\bf b}, $E_F=20$ meV.
In {\bf c}, the lower plasmon frequency is calculated and compared with approximated expressions.
The exact value is obtained by numerically solving ${\rm Re}[n^2_{\pm}(\omega_{\pm})]=0$.
Equation~\eqref{eq:plasma-frequency} is very close to the exact value.
The approximation $\omega_{p,\rm lower}\approx D/\sigma_H(0)$ becomes more accurate when $\sigma_H$ is larger.
{\bf d}-{\bf e}, Electron-doped FeSe model with chiral superconducting pairing $\Delta_0=10$ meV.
$\Gamma=0.1$ meV.
The reflectances and Kerr angles are calculated for the normal incidence on bulk crystals.
We take the interlayer lattice constant $a_z=5$ \AA for all calculations.
}
\label{fig:MOKE}
\end{figure*}

{\bf \noindent Magneto-optic Kerr effects.}
Let us explicitly calculate the polar magneto-optic Kerr effects resulting from the optical Hall conductivity in our models.
At the interface between the vacuum and the medium, the reflection coefficient matrix $r$ relates the electric fields of reflected and incident lights by ${\bf E}_r=r{\bf E}_i$, and the transmission coefficient is given by $t=1-r$ (this is transmission to the bulk, not through).
Since the circular polarization of light is preserved under reflection at normal incidence by the $C_{4z}$ symmetry in our models, we have $r_{+-}=r_{-+}=0$, so the circular-polarization dependence is encoded in the difference between $r_{++}=(1-n_{+})/(1+n_{+})$ and $r_{--}=(1-n_{-})/(1+n_{-})$.

We consider the Dirac model with $\Delta_y=10$ meV and the electron-doped FeSe model with $\Delta_0=10$ meV.
In both models, at frequencies below the lower plasmon frequency $\omega_{p,-}$, perfect reflection occurs for all polarizations (i.e., $R_+\equiv |r_{++}|^2=1$ and $R_-\equiv |r_{--}|^2=1$) because of the absence of transmission and absorption [Figs.~\ref{fig:MOKE}a,d].
$\omega_{p,-}$ is smaller for $M>0$ than $M<0$ in the Dirac model due to the stronger Hall effect in the former [Figs.~\ref{fig:MOKE}c].
The Kerr angle $\phi_K\equiv \tan^{-1}(r_{yx}/r_{xx})=\tan^{-1}[i(r_{++}-r_{--})/(r_{++}+r_{--})]$ is purely real-valued in this frequency regime, and its size increases monotonically [Figs.~\ref{fig:MOKE}b,e].
Above $\omega_{p,-}$, circular-polarization-selective perfect reflection occurs up to the optical excitation gap $E_g/\hbar$, and the Kerr angle develops a finite imaginary part.
\\

{\bf \noindent Discussion.}
Our classification based on the selection rule provides a unified understanding of the significant Hall effect in chiral superconductors.
Although we focused on superconductivity from a chiral metal as an example that employs $PC$-asymmetric channels, we can also make these channels significant by considering chiral superconductivity in systems with (either exactly or approximatly) degenerate Fermi surfaces.
Since the $PC$ symmetry forbids only one transition channel for a given Bloch state, multiple surviving channels can exist at the pairing scale when there is degeneracy.
This kind of example was provided in spin-orbit coupled superconductors with nonunitary pairing~\cite{zhang2024quantum}.
We also note that the large Hall effect in models of the twisted cuprates~\cite{can2021probing,song2022doping} can be understood as resulting from inversion symmetry breaking.
Therefore, our criteria in Fig.~\ref{fig:selection-rule} provide a unified explanation of the known mechanisms of the significant Hall effect in superconductors.
We remark that large magneto-optic effects in superconductors provide an exciting possibility of controlling the chiral superconductivity using optical methods.
Our results will provide a useful guide for research in this direction.

While we focus on the new phenomenon of the circular-polarization-selective perfect reflection, many proposals for a chiral cavity are based on the optical rotation~\cite{hubener2021engineering}.
In fact, the Kerr-rotating perfect reflection, occurring universally in gapped chiral superconductors, can also be used for a chiral cavity.
In this case, the relative strengths of two circularly polarized fields will change spatially, so it is difficult to couple matter with a specific sign of chirality if it fills the whole cavity.
On the other hand, the spatial variation can be a benefit in that the chirality is tunable by changing the location of matter inside the cavity, for the purpose of engineering a thin two-dimensional material or small molecules.
Therefore, depending on the application, the Kerr-rotating perfect reflection can be more appropriate than the circular-polarization-selective perfect reflection, and vice versa.
In any case, it is desired to realize circular-polarization-selective superconducting mirrors because they can also show giant Kerr rotations near the plasma frequencies.

Our model analysis above assumes clean superconductors where optical excitations, either virtual or real, preserve crystal momentum.
Let us discuss the effect of impurities.
In our model calculations for the superconductivity from a chiral metal, the role of superconductivity is minor because it changes the optical properties, such as the response functions and the excitation gap, only slightly.
Therefore, one might have an impression that the same phenomenon described here occurs in the normal metallic state as well.
However, this is an artifact of being in the clean limit.
In practice, superconducting states  are advantageous in terms of their robustness against Ohmic losses~\cite{raimond2001manipulating}.
In the presence of a nonzero density of impurities, the Drude absorption occurs at any finite frequency in normal metals:
the absorptance $A_{\pm}(\omega)=O(\gamma/\omega_{p,\pm})$ is finite even at low frequencies $\omega<\omega_{p,\pm}$ due to impurity scattering, where $\gamma$ is the current relaxation rate.
The Drude absorption becomes particularly significant when the plasmon frequency is small --- which is required to achieve circular-polarization-selective perfect reflection at low frequencies.
This is in contrast to the case of superconductors, that show vanishing in-gap optical absorption.
Because the Drude absorption involves finite momentum transfer, which in general is not $PC$ symmetric, it is not prohibited by symmetry around $E_g$~\cite{mattis1958theory}.
So, the optical excitation gap reduces to the superconducting gap, and no excitation is allowed only below this gap.
Accordingly, the circular-polarization-dependent perfect reflection for $M<0$ in Fig.~\ref{fig:MOKE}a, which appears for $\hbar\omega>E_g$, will disappear for weak impurities, while that for $M>0$ remains robust as long as the impurity scattering rate is smaller than the difference between $\omega_{p,-}$ and $E_g/\hbar$.
One aspect to note is that, while Anderson's theorem~\cite{anderson1959theory} guarantees that the superconducting gap is unaffected by weak impurities for conventional $s$-wave pairing, weak impurities can influence the gap size for general unconventional pairing~\cite{andersen2020generalized,timmons2020electron,cavanagh2021general}.

For material realizations, the criterion in Eq.~\eqref{eq:bound} is quite difficult to satisfy.
However, a candidate material already exists.
A recent experiment~\cite{han2024signatures} discovered signatures of chiral superconducting phases in doped rhombohedral tetra- and penta-layer graphene.
There, superconductivity emerges in the vicinity of a chiral metal, showing significant Hall conductivity in the order of 2$e^2/h$ and a small Fermi energy of a few meV.
Moreover, the superconductivity is reported to be in the BCS-BEC crossover regime~\cite{han2024signatures}, showing a comparable Fermi energy and  pairing gap.
Thus, it is likely that Eq.~\eqref{eq:bound} is satisfied.
It should be noted, however, that rhombohedral graphene is a two-dimensional material, while for perfect reflection at normal incidence, we need a three-dimensional material with sufficient thickness, that exceeds  the light penetration depth.
Therefore, to observe circular-polarization-selective perfect reflection in doped rhombohedral graphene at normal incidence, we need to stack them three-dimensionally while preserving their superconducting characters, which may be challenging to realize.
Nonetheless, even in its two-dimensional form, doped rhombohedral graphene may exhibit notable magneto-optic effects.
In particular, when Eq.~\eqref{eq:bound} is satisfied, a significant Kerr effect is expected within the pairing gap, presenting opportunities for further exploration.

To put our work in a broader scope, let us note that the role of topology~\cite{de2017quantized,alexandradinata2022topological,jankowski2024quantized,onishi2024fundamental} and, more broadly, geometry~\cite{morimoto2016topological,holder2020consequences,watanabe2021chiral,ahn2020low,ahn2022riemannian} of quantum states in functional optical properties of materials has been intensively studied recently.
In most cases, optical responses are not quantized, so they are not topological in the strict sense.
However, even for non-quantized responses, nontrivial topology often indicates a larger response, as in our model calculations.
It remains an interesting open question to formulate rigorous topological bounds on the optical responses of superconductors.

Finally, we conclude with a remark that the implications of our findings extend beyond fundamental quantum materials research into other applications. Since mirrors are fundamental elements in optics, our result will find broad applications in chiral photonics, including the design of circularly polarized lasers, and pave the way for innovative technologies.
\\

{\bf \large \noindent Acknowledgements}\\
{\small We appreciate Ceren B. Dag, Junkai Dong, Patrick Ledwidth, Naoto Nagaosa, Edoardo Baldini, and Chandra Varma for helpful discussions.
This research was supported by the Center for Advancement of Topological Semimetals, an Energy Frontier Research Center funded by the U.S. Department of Energy Ofﬁce of Science, Ofﬁce of Basic Energy Sciences, through the Ames Laboratory under Contract No. DE-AC02-07CH11358.}\\

{\bf \noindent \large Methods}\\
{\bf \noindent Optical conductivity formulas.}
The optical conductivity of fermions in the Bogoliubov-de Gennes formalism is given by
\begin{align}
\sigma^{ab}(\omega)
&=\frac{i}{2}
\int_{\bf k}\frac{1}{\omega}\sum_{n}f_{n}K^{ab}_{nn}
+\frac{1}{\hbar}\sum_{n,m}\frac{f_{nm}J^a_{nm}J^b_{mn}}{(\omega-\omega_{mn})\omega},
\end{align}
where $K^{ab}_{mn}=\braket{\psi_m|\d_{A_a}\d_{A_b}\hat{H}|\psi_n}|_{{\bf A}=0}$, $J^{a}_{mn}=\braket{\psi_m|\d_{A_b}\hat{H}|\psi_n}|_{{\bf A}=0}$, and $\ket{\psi_n}$ is the Bloch energy eigenstate with energy $\hbar\omega_n$.
The overall factor $1/2$ appears because we consider single-particle eigenstates in the Nambu spinor basis, which contain doubled degrees of freedom.
This factor does not appear in formulas based on the physical electronic degrees of freedom [see e.g., the formula in ref.~\cite{parker2019diagrammatic} for non-superconducting states].
The above formula can be rewritten as
\begin{align}
\label{eq:conductivity2}
\sigma^{ab}(\omega)
&=\frac{i}{\omega}D^{ab}
+\frac{i}{2\hbar}\int_{\bf k}\sum_{n,m}\frac{f_{nm}J^a_{nm}J^b_{mn}}{(\omega-\omega_{mn})\omega_{mn}},
\end{align}
where
\begin{align}
\label{eq:SF-weight}
D^{ab}
&=\frac{\hbar}{2}
\int_{\bf k}\sum_{n}f_{n}\d_{A_a}\d_{A_b}\omega_n|_{{\bf A}=0}\notag\\
&=\frac{1}{2}
\int_{\bf k}
\sum_{n}f_{n}K^{ab}_{nn}
-\sum_{n,m}f_{nm}\frac{J^a_{nm}J^b_{mn}}{\hbar\omega_{mn}}
\end{align}
is the superfluid weight defined by $j^a(0)=-D^{ab}A^b(0)$.
The expression in the second line of Eq.~\eqref{eq:SF-weight} corresponds to the generalization of the Ferrel-Glover-Tinkham (FGT) sum rule to low-energy models where $K^{ab}_{nn}\ne \delta_{ab}e^2/m_e$, and $m_e$ is the bare electron mass.
To derive Eq.~\eqref{eq:conductivity2}, we use
$(\d_{A_a}\hat{O})_{mn}
\equiv \braket{\psi_m|\d_{A_a}\hat{O}|\psi_n}
=\d_{A_a}O_{mn}-i[\alpha^a,O]_{mn}$,
where ${\bm \alpha}=\braket{\psi_m|i\d_{\bf A}\psi_n}$.
In particular, $\alpha^a_{mn}=-J^a_{mn}(i\hbar\omega_{mn})^{-1}$ for $m\ne n$ because
$J^a_{mn}
=-\hbar(\delta_{mn}\d_{A_a}\omega_{n}
+i\omega_{mn}\alpha^a_{mn})$.

The symmetric and Hall parts of the conductivity tensor have the form
\begin{align}
\sigma^{ab}_S(\omega)
&\equiv \frac{\sigma^{ab}+\sigma^{ba}}{2}\notag\\
&=\frac{i}{\omega}D^{ab}
-i\omega \epsilon_0\tilde{\chi}_S^{ab}(\omega),\notag\\
\tilde{\chi}_S^{ab}(\omega)
&=\frac{1}{2\epsilon_0\hbar}
\int_{\bf k}\sum_{n\in {\rm occ},m\in {\rm unocc}}
\frac{J^a_{nm}J^b_{mn}+J^b_{nm}J^a_{mn}}{\omega_{mn}(\omega^2_{mn}-\omega^2)},\notag\\
\sigma^{ab}_H(\omega)
&\equiv \frac{\sigma^{ab}-\sigma^{ba}}{2}\notag\\
&=\frac{i}{2\hbar}
\int_{\bf k}\sum_{n\in {\rm occ},m\in {\rm unocc}}
\frac{J^a_{nm}J^b_{mn}-J^b_{nm}J^a_{mn}}{(\omega^2-\omega^2_{mn})}.
\end{align}

The matrix elements of $\hat{\cal H}^{a_1\hdots a_n}\equiv \d_{A_{a_1}}\hdots \d_{A_{a_n}}\hat{H}$ can be calculated from the single-particle Bogoliubov-de Gennes Hamiltonian:
$\braket{\psi_{m\bf k}|\hat{\cal H}^{a_1\hdots a_n}|\psi_{n\bf k}}
=\braket{u_{m\bf k}|u_{\mu \bf k}}[\d_{A_{a_1}}\hdots \d_{A_{a_n}}H_{\rm BdG}({\bf k})]_{\mu\nu}\braket{u_{\nu \bf k}|u_{n\bf k}}$
where
$\ket{\psi_{n\bf k}}=e^{i{\bf k}\cdot {\bf r}}\ket{u_{n\bf k}}$ and $\hat{H}({\bf k})=e^{-i{\bf k}\cdot {\bf r}}\hat{H}e^{i{\bf k}\cdot {\bf r}}$.
We consider light-matter interaction through the minimal coupling ${\bf k}\rightarrow {\bf k}-q{\bf A}/\hbar$ in the kinetic part $\hat{c}^{\dagger}h\hat{c}$.
Since the electric charge $q=-e$ ($+e$) for the electron $\hat{c}$ (hole $\hat{c}^{\dagger}$) sector,
\begin{align}
{\cal H}^{a_1\hdots a_n}
\equiv \d_{A_{a_1}}\hdots \d_{A_{a_n}}H_{\rm BdG}
=\left(\frac{e}{\hbar}\tau_z\right)^n\d_{a_1}\hdots\d_{a_n}H_{\rm BdG},
\end{align}
where we use the notation $\d_a=\d_{k_a}$. At the first and second orders, we have
\begin{align}
\label{eq:J-K}
J^a({\bf k})
&\equiv -{\cal H}^{a}({\bf k})
=-\frac{\d H_{\rm BdG}}{\d A_a}\bigg|_{{\bf A}=0}
=-\frac{e}{\hbar}\tau_zH_{\rm BdG}({\bf k}),\notag\\
K^{ab}({\bf k})
&\equiv {\cal H}^{ab}({\bf k})
=\frac{\d^2 H_{\rm BdG}}{\d A_a\d A_b}\bigg|_{{\bf A}=0}
=\frac{e^2}{\hbar^2}\d_{a}\d_{b}H_{\rm BdG}({\bf k}).
\end{align}
When the pairing originates from a nonlocal electron interaction, the pairing function is also coupled to the gauge field, as shown in Ref.~\cite{oh2024revisiting}, so $J$ and $K$ do not have the form in Eq.~\eqref{eq:J-K} and instead depend on the type of interactions.
We do not consider this complication by assuming that pairing is due to local interactions.
Our selection rule in Eq.~\eqref{eq:selection-rule} is not affected by the presence of nonlocal interactions.\\

{\bf \noindent Derivation of equation~\eqref{eq:bound}.}
Let us define $E_C\equiv e^2/2\pi \tilde{\epsilon}a_z$ and $c_0\equiv \sigma_H(0)a_zh/e^2$.
Equation~\eqref{eq:plasma-frequency} is then
\begin{align}
\hbar\omega_{p,\pm}
&\approx E_C\left(\sqrt{\hbar^2D/(\tilde{\epsilon}E_C^2)+(c_0/2)^2}\pm c_0/2\right).
\end{align}
We have $\hbar\omega_{p,\rm lower}=E_g$ when $D=D_c$, where
\begin{align}
D_c\equiv \frac{\tilde{\epsilon}}{\hbar^2}(E_g^2+E_gE_C|c_0|).
\end{align}
Since $\hbar\omega_{p,\rm lower}$ increases monotonically with $D\ge 0$, the inequality $\hbar\omega_{p,\rm lower}<E_g$ is equivalent to $D<D_c$.
For a cylindrical Fermi surface with $n_s$ degeneracy, we obtain
\begin{align}
\frac{E_g}{E_F}
>\frac{n_s}{E_g/E_C+|c_0|}
\end{align}
by using $D=n_sE_FE_C\tilde{\epsilon}/\hbar$.\\

{\bf \noindent Tight-binding model of electron-doped iron selenide.}
We take the local $d$-orbital basis of Ref.~\cite{eschrig2009tight}
\begin{align}
(d_{xy},d_{x^2-y^2},\pm id_{yz},\pm id_{xz},d_{3z^2-r^2})
\end{align}
for two sublattices $\pm$.
The pure imaginary factor in front of $yz$ and $xz$ orbitals ensures that all basis states are invariant under $C_{2z}T$, such that all the single-particle Hamiltonian matrix elements are real-valued in two-dimensional momentum space.
The choice of the alternating sign makes the orbital matrix representation of the glide mirror operator $G_z:(x,y,z)\rightarrow (x+1/2,y,-z)$ simple: $\braket{d_{i\alpha}|\hat{M}|d_{j\beta}}=\delta_{\alpha\beta}$, where the sites $i$ and $j$ are related to each other by the glide mirror operation (${\bf r}_i=G_z{\bf r}_j$).

In this orbital basis, we consider the tight-binding model
\begin{align}
\hat{H}
=\sum_{i,\alpha,j,\beta}\tilde{Z}_{\alpha\beta}t_{i\alpha j\beta}\hat{c}^{\dagger}_{i\alpha}\hat{c}_{j\beta}
+\frac{1}{2}(\Delta_{i\alpha j\beta}\hat{c}^{\dagger}_{i\alpha}\hat{c}^{\dagger}_{j\beta}+h.c.)
\end{align}
where $t$ is the bare hopping parameter, $\tilde{Z}_{\alpha\beta}\equiv\sqrt{Z_{\alpha}Z_{\beta}}$, and $Z$ is the renormalized quasiparticle weight.
We take the hopping parameters and onsite energies from the ab initio calculations for pristine two-dimensional FeSe in Ref.~\cite{eschrig2009tight}.
To produce the band structure that mimics that of the heavily electron-doped FeSe, we include correlation effects by taking $Z=(0.104784, 0.1977, 0.282881, 0.282881, 0.787272)$, and, additionally, rescale the intra-orbital nearest-neighbor hopping amplitudes for the $d_{xy}$ orbital and $d_{xz,yz}$ orbitals by $1.3$ and $3$, respectively, and shift the onsite energies by $(50,50,-85,-85,50)$ meV.\\

\bibliographystyle{naturemag}

\end{document}